\documentclass[aps,prd,preprint,nofootinbib,superscriptaddress,tightenlines]{revtex4}
\usepackage{amsfonts}
\usepackage{mathrsfs}
\usepackage{graphicx}
\usepackage{amsmath}
\usepackage{amssymb}
\usepackage{multirow}
\usepackage{subfigure}
\usepackage{epsfig}
\usepackage{graphicx}
\usepackage{booktabs}
\usepackage{array}
\usepackage{tabularx}
\usepackage{slashed}
\usepackage{ulem}
\usepackage{bm}
\usepackage{hyperref}
\usepackage{color}

\newcommand{\bqa}{\begin{eqnarray}}
\newcommand{\eqa}{\end{eqnarray}}
\newcommand{\beq}{\begin{equation}}
\newcommand{\eeq}{\end{equation}}
\graphicspath{{fig/}{dia/}} \DeclareGraphicsExtensions{.eps}
\begin{document}
\baselineskip 20pt
\title{Exclusive production of doubly charmed baryon pair at B Factories}

\author{Xiao Liang}
\email{lx321@sdut.edu.cn}
\affiliation{School of Physics and Optoelectronic Engineering, Shandong University of Technology, Zibo, Shandong 255000, China}

\author{Jun Jiang}
\email{jiangjun87@sdu.edu.cn}
\affiliation{School of Physics, Shandong University, Jinan, Shandong 250100, China}

\author{Chun-Yuan Li}
\email{lichunyuan@sdut.edu.cn}
\affiliation{School of Physics and Optoelectronic Engineering, Shandong University of Technology, Zibo, Shandong 255000, China}

\author{Meng-Li Liu}
\email{26412011158@stumail.sdut.edu.cn}
\affiliation{School of Physics and Optoelectronic Engineering, Shandong University of Technology, Zibo, Shandong 255000, China}

\author{Xing-Hua Yang}
\email{yangxinghua@sdut.edu.cn}
\affiliation{School of Physics and Optoelectronic Engineering, Shandong University of Technology, Zibo, Shandong 255000, China}

\author{Zong-Guo Si}
\email{zgsi@sdu.edu.cn}
\affiliation{School of Physics, Shandong University, Jinan, Shandong 250100, China}

\begin{abstract}

The exclusive production of doubly charmed baryon pairs $\Xi_{cc}+\bar{\Xi}_{cc}$ via $e^+e^-$ annihilation at B factories are calculated in the framework of non-relativistic QCD (NRQCD). 
The total cross sections, theoretical uncertainties, differential distributions and event yields for the BaBar and Belle II experiments are systematically discussed. 
According to our study, the detection prospects for both experiments are unfavorable.

\vspace {2mm} 
\end{abstract}

\maketitle

\section{INTRODUCTION}
\label{sec:introduction}

Doubly charmed baryons are composed of two charm quarks and one light quark. Their production consists of perturbative production of constituent quarks and nonperturbative hadronization into the baryon, which provides favorable conditions for theoretical studies of strong interactions at both perturbative and nonperturbative levels.
All along, high-energy collider experiments have been continuously searching for doubly heavy baryons, and the evidence of its existence was found as early as 2002 by SELEX Collaboration \cite{SELEX:2002wqn}, but it is also pointed out that this evidence may lack sufficient support \cite{Kiselev:2002an,Ratti:2003ez,BaBar:2006bab,Belle:2006edu,LHCb:2013hvt}. In 2017, theoretical studies regarding the weak decays of doubly heavy baryons predicted that the doubly charmed baryons $\Xi_{cc}^{++}$ are most likely to be discovered through the two decay channels $\Xi_{cc}^{++} \to \Lambda_c^+K^-\pi^+\pi^+$ and $\Xi_{cc}^{++} \to \Xi_{c}^{+}\pi^+$ \cite{Yu:2017zst}. Subsequently, LHCb collaboration reported experimental evidence for the $\Xi_{cc}^{++}(3621)$ signal founded in the $\Lambda_c^+K^-\pi^+\pi^+$ invariant mass \cite{LHCb:2017iph} in 2017. Then the LHCb collaboration further confirmed the existence of the doubly charmed baryon $\Xi_{cc}^{++}$ via the decay channel $\Xi_{cc}^{++} \to \Xi_{c}^{+}\pi^+$ \cite{LHCb:2018pcs}. Its lifetime, mass and production cross section were also precisely measured by the LHCb \cite{LHCb:2018zpl,LHCb:2019epo,LHCb:2019qed}, which also uncovered a new decay channel  $\Xi_{cc}^{++} \to \Xi_{c}^{0}\pi^+\pi^+$ \cite{LHCb:2025shu}.
In addition, the LHCb collaboration reported the observation of the $\Xi_{cc}^{+}$ baryon via the decay mode $\Xi_{cc}^{+} \to \Lambda_c^+K^-\pi^+$, with a statistical significance above $7\sigma$ \cite{LHCb:2026pxn}.
Furthermore, the latest LHCb measurement has confirmed the existence of $\Omega_{cc}^+$ \cite{LHC}.

Considering the fact that the constituent doubly charmed quark pair inside the doubly charmed baryon has the nonrelativistic nature, we adopt the nonrelativistic QCD (NRQCD) factorization formalism proposed by Bodwin, Braaten and Lepage \cite{Bodwin:1994jh,Petrelli:1997ge} to study the doubly charmed baryon
production.
This framework was originally developed for calculations of quarkonium \cite{Bodwin:1994jh,Petrelli:1997ge,Bodwin:2002cfe}. It provides a systematic way to factor the process into the perturbative creation of constituent heavy quarks and the nonperturbative hadronization. 
Within NRQCD, the baryon productions from electron-positron annihilation represents a major theoretical challenge. 
For example, the experimental results \cite{Belle:2002tfa,Pakhlov:2004au,BaBar:2005nic} for the exclusive production process $e^+e^-\to J/\psi+\eta_c$ at B factories are larger than leading order (LO) theoretical predictions \cite{Braaten:2002fi,Liu:2002wq} by at least a factor of 5 up to an order of magnitude. 
This large discrepancy between theory and experiment finally be resolved by the next-to-leading-order (NLO) corrections within NRQCD \cite{Zhang:2005cha}.
The NRQCD factorization framework was later extended to studies of doubly heavy baryons \cite{Jiang:2012jt,Jiang:2013ej}.

To date, the production of doubly heavy baryons has been widely studied, with numerous groups conducting calculations for inclusive processes involving their excited states \cite{Ma:2003zk,Yang:2014ita,Yang:2024ysg,Niu:2023ojf,Ma:2022ger,Jiang:2012jt,Jiang:2013ej,Chen:2014frw,Bi:2017nzv,Sun:2020mvl,Chang:2006eu,Chang:2006xp,Zhang:2022jst,Luo:2022lcj,Niu:2019xuq,Ma:2022cgt,Shim:2025dgq,Zhao:2025kvq,Ma:2025ito,Tian:2023uxe}. Among which, there are extensively studies for the inclusive production of doubly heavy baryon through $e^+e^-$ collision \cite{Ma:2003zk,Yang:2014ita,Niu:2023ojf,Jiang:2012jt,Jiang:2013ej,Chen:2014frw,Ma:2025ito}, $ep$ collision \cite{Bi:2017nzv,Sun:2020mvl}, $pp$ collision \cite{Chang:2006eu,Chang:2006xp} and ultraperipheral collision \cite{Yang:2024ysg,Zhao:2025kvq}, as well as the indirect productions via $W$, $Z$, Higgs and top quark \cite{Ma:2022ger,Zhang:2022jst,Luo:2022lcj,Niu:2019xuq,Ma:2022cgt,Tian:2023uxe}. The above studies may offer chances for experimental research on doubly heavy hadrons and create a new testing platform for the NRQCD factorization formalism in different collider environments.
In this work, we have focused our attention on calculating the exclusive production of doubly charmed baryon pair via electron-positron collision at B factories like BaBar and Belle. 

The rest of the paper is organized as follows. 
In section \ref{sec:formulation},  
the main theoretical framework of the exclusive production of doubly charmed baryon is presented.
In section \ref{sec:data}, comprehensive numerical results will be discussed.
Sec.\ref{sec:summary} is reserved for a summary.

\section{FORMULATION}
\label{sec:formulation}
\begin{figure}[!thbp]
    \centering   \includegraphics[width=0.8\textwidth]{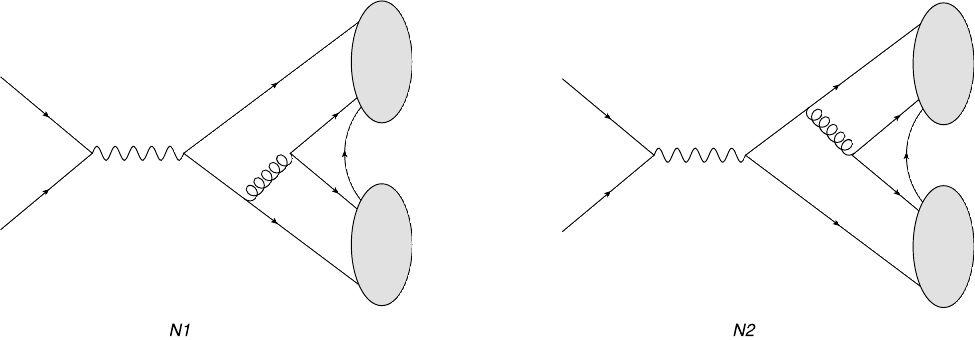}
    \caption{Typical Feynman diagrams for $e^{+}+e^{-}\to \gamma^{*} \to \Xi_{cc}+\bar{\Xi}_{cc} $.}
    \label{diagram}
\end{figure}

We investigate the process $e^{+}+e^{-} \to \Xi_{cc}+\bar{\Xi}_{cc}$, where $\Xi_{cc}$ denotes the doubly charmed baryon and $\bar{\Xi}_{cc}$ refers to the corresponding doubly charmed antibaryon.
The typical tree-level Feynman diagrams for $e^{+}+e^{-} \to \Xi_{cc}+\bar{\Xi}_{cc}$ are displayed in Fig. \ref{diagram}. As illustrated therein, the collision of an electron–positron pair first produces a $cc$ diquark and a $\bar{c}\bar{c}$ anti-diquark. By absorbing one constituent from a light quark–antiquark pair generated from the vacuum, they form a doubly charmed baryon and its corresponding antibaryon, respectively.
Since we specialize to the production in the B factory, which means that the dominant contribution comes from $\gamma$ mediation, and the production of Z-boson mediation is ignored.

Within the NRQCD framework, the differential cross section of the process $e^+(p_a)+e^-(p_b)\to \Xi_{cc}(p_1)+\bar{\Xi}_{cc}(p_2)$ can be factored into the short-distance coefficients and the long-distance matrix elements,
\begin{equation}
d\sigma(e^+e^- \to \Xi_{cc}\bar{\Xi}_{cc})= d\hat\sigma(e^+e^- \to (cc)[n]+(\bar{c}\bar{c})[n'])    {\langle  \mathcal{O}^H(n)  \rangle}   {\langle  \mathcal{O}^H(n')  \rangle}  \  .
\label{dsigma}
\end{equation}
The short-distance coefficients $\hat\sigma$ describe the short-distance production of intermediate Fock states with definite spin and color quantum numbers. The long-distance non-perturbative matrix elements ${\langle  \mathcal{O}^H(n)  \rangle}$ and ${\langle  \mathcal{O}^H(n')  \rangle}$ describe the hadronization of the Fock state into the doubly charmed baryon $\Xi_{cc}$ and antibaryon $\bar{\Xi}_{cc}$, which can be related to the Schr\"odinger wavefunctions at the origin \cite{Petrelli:1997ge,Quigg:1977dd,Braaten:1993jn} ${\langle  \mathcal{O}^H(n)  \rangle}={\langle  \mathcal{O}^H(n')  \rangle} = |\Psi_{\Xi_{cc}}(0)|^2$. 

The short-distance differential cross section $d\hat\sigma$ is perturbatively calculable, and can be expressed as
\begin{eqnarray}
\begin{split}
d\hat\sigma= 
&\frac{1}{2s} \overline{\sum}  \big|{\cal M}\big|^{2} d\Phi_2 \ , 
\end{split}
\label{sd-sigma}
\end{eqnarray}
where $\overline{\sum}$ stands for the average over the spin of the initial electron and positron, and sum over the color and spin of the final particles when manipulating the squared amplitudes. $d\Phi_2$ stands for the two-body phase space of final states.

The diquark production scattering amplitude can be obtained through an appropriate transformation of the related meson production scattering amplitude. Then we can derive the diquark amplitude in the same way as calculating meson production \cite{Jiang:2012jt}. In the case of the process $e^+(p_a)+e^-(p_b)\to (cc)[n](p_1)+(\bar{c}\bar{c})[n](p_2)$, the scattering amplitude can be expressed as,
\begin{equation}
    \begin{split}
    {\cal M} 
      = & \ \bar{u}(\boldsymbol{p}_{11}) \, . \, x \, . \, v(\boldsymbol{p}_{22})\times \bar{u}(\boldsymbol{p}_{12}) \, . \, y \, . \, v(\boldsymbol{p}_{21}) 
      \times {\cal C}_{M_{\Xi_{cc}}} \times {\cal C}_{M_{\bar{\Xi}_{cc}}} \times {\cal D}  \times  {\cal L} \ ,
 \end{split}
    \label{diquarkline}
\end{equation}
where Lorentz indices are all omitted. The $p_1$ and $p_2$ are the momenta of the doubly charmed baryon and doubly charmed antibaryon, respectively. $p_{11}$ and $p_{12}$ denote the momenta of the two constituent charm quarks inside the doubly charmed baryon. Similarly, $p_{21}$ and $p_{22}$ are the momenta of the two constituent anti-charm quarks inside the doubly charmed antibaryon. At the leading velocity expanding in NRQCD they have the relationship $p_{11}=p_{12}=p_1/2$ and $p_{21}=p_{22}=p_2/2$. The $x$ and $y$ represent the interaction vertexes, fermion propagator, and gluon propagator along the corresponding spinor lines. $\cal C$ are the color factors of baryons. ${\cal D}$ serves as the photon propagator in this context, and ${\cal L}$ represents the leptonic part, they can be written as 
\begin{equation}
    \begin{split}
    &{\cal D}=   \frac{-i g_{\mu\nu}}{s}, \\
    &{\cal L} = \bar{v}(p_{a}) (i e\gamma^\mu )  u(p_{b}) .
 \end{split}
\end{equation}

Then we can transform the Dirac fermion chain of the diquark production scattering amplitude in equation (\ref{diquarkline}) with a charge conjugation $C$,
\begin{eqnarray}
&& v^T C = -\bar{u},\;\;\;C^- \Gamma^T_i C = -\Gamma_i,\;\;\;CC^- = 1, \\
&& C^-s^T_f(k,m)C = s^T_f(-k,m),\;\;\; C^-\bar{u}^T = v.
\end{eqnarray}
Here, $\Gamma_i$ represents the interaction vertex (the $\gamma$-matrix elements), and $s_f(k,m)$ is the fermion propagator in between the interaction vertices. Therefor, diquark production scattering amplitude in equation (\ref{diquarkline}) becomes,
\begin{equation}
    \begin{split}
    {\cal M} 
      = (-1)^{(\rho+1)} \times \bar{u}(\boldsymbol{p}_{11}) \, . \, x \, . \, v(\boldsymbol{p}_{22})\times \bar{u}(\boldsymbol{p}_{21}) \, . \, y' \, . \, v(\boldsymbol{p}_{12}) \times {\cal C}_{M_{\Xi_{cc}}} \times {\cal C}_{M_{\bar{\Xi}_{cc}}}\times {\cal D}  \times  {\cal L} \ ,
 \end{split}
\end{equation}
where $\rho$ is the number of interaction vertex in the fermion line $y'$. It shows that the diquark production is related to the quarkonium production by a simple factor $(-1)^{(\rho+1)}$. Then we proceed with the calculation using projection operators.
\begin{equation}
    \begin{split}
    {\cal M}
      = (-1)^{(\rho+1)} \times Tr[ \Pi_{\Xi_{cc}}(p_{1}) \, . \, x \, . \, \Pi_{\bar{\Xi}_{cc}}(p_{2})  \, . \, y' ] \times {\cal C}_{M_{\Xi_{cc}}} \times {\cal C}_{M_{\bar{\Xi}_{cc}}} \times {\cal D}  \times  {\cal L} \ .
 \end{split}
    \label{mesonline}
\end{equation}
Where $\Pi_{\Xi_{cc}}$ and $\Pi_{\bar{\Xi}_{cc}}$ denote the projection operators of doubly charmed baryon $\Xi_{cc}$ and its antibaryon counterpart $\bar{\Xi}_{cc}$, respectively. They have the form
\begin{equation}
\begin{split}
      \Pi_{\Xi_{cc}}(p_1)= \frac{1}{2\sqrt{M_{\Xi_{cc}}}} \slashed{\epsilon}(p_1) (\slashed{p}_{1} + M_{\Xi_{cc}}) \ , \\  \Pi_{\bar{\Xi}_{cc}}(p_2)= \frac{1}{2\sqrt{M_{\bar{\Xi}_{cc}}}} \slashed{\epsilon}(p_2) (\slashed{p}_{2} + M_{\bar{\Xi}_{cc}}) \ .
       \label{eq:projection}
       \end{split}
\end{equation}
Here $\epsilon(p_1)$ and $\epsilon(p_2)$ are the polarization vector of $(cc)$-diquark and $(\bar{c}\bar{c})$-antidiquark, respectively. In this paper, we only consider the vector intermediate Fock state $[^3S_1]$.


In the production process of the doubly charmed baryon, the $(cc)$ di-quark is initially formed in color state $\bar{3}$ and then binds to the light quark in color state $3$,  while the $(\bar{c}\bar{c})$ anti-diquark formed in color state $3$ and binds to the anti-light quark in color state $\bar{3}$. The baryon and antibaryon formed in these configurations must be color neutral due to the property of color confinement.
Therefore the color factor ${\cal C}_{M_{\Xi_{cc}}}$ and ${\cal C}_{M_{\bar{\Xi}_{cc}}} $ in Eq. (\ref{mesonline}) utilize the anti-symmetric function, so their product ${\cal C}_{M_{\Xi_{cc}}} \times {\cal C}_{M_{\bar{\Xi}_{cc}}}$ has the form
\begin{eqnarray}
{\cal C}_{M_{\Xi_{cc}}} \times {\cal C}_{M_{\bar{\Xi}_{cc}}} = N \times \varepsilon_{ijk} \times \varepsilon_{nm\bar{k}} \times \delta_{k \bar{k}} = N \times (\delta_{in}\delta_{jm}-\delta_{im}\delta_{jn}) \ .
\label{color}
\end{eqnarray}
The indices $i, j$ and $k$ are color indices of each constituent charm quark and the diquark $(cc)[n]$, respectively. The indices $n, m$ and $\bar{k}$ are color indices of each constituent charm anti-quark and the anti-diquark $(\bar{c}\bar{c})[n]$, respectively. $\delta_{k\bar{k}}$ originates from the light quark pair, and the normalization constant $N=1/\sqrt{12}$. 

From the above formulation, we finally obtain the cross sections for the exclusive production of doubly charmed baryons at B factories as follows,
\begin{eqnarray}
 \begin{split}
\sigma  \ =  \
 \frac{1024 \, |\Psi_{\Xi_{cc}}(0)|^4 \, \pi^3 \, \alpha_s^2 \, \alpha^2 \, C_A^2 \, C_F^2 \, (s-16\,M_{\Xi_{cc}})^{3/2} \, (192\,M_{\Xi_{cc}}^2+ 80\, M_{\Xi_{cc}} \, s+s^2)  }{81 \, M_{\Xi_{cc}} \, s^{13/2} } \ . 
  \end{split}
  \label{sigma}
\end{eqnarray}
Here, the symmetry factors arising from identical constituent charm quarks in the doubly charmed baryon and identical anti-charm quarks in the doubly charmed antibaryon have already been taken into account.
\section{NUMERICAL RESULTS}
\label{sec:data}

\subsection{Total cross sections}
\label{Total cross sections}
When doing the numerical calculation, the necessary input parameters are
taken as $\alpha=1/137.065$ and $  \alpha_{s}(2m_c)=0.242 $ \cite{ParticleDataGroup:2024cfk}. For the squared wave functions at the origin of doubly charmed baryons $|\Psi_{\Xi_{cc}}(0)|^2$, we employ the squared wave functions at the origin of $[1S]$ state charmed diquark $|\Psi_{cc[1S]}(0)|^2=0.0390$ \cite{Baranov:1995rc} as an estimate. This quantity is calculated under the Power-low potential.

By taking the center-of-mass (CM) energy of B factory to be $\sqrt{s} = 10.58 \; {\rm GeV}$, we present the LO total cross section for $e^++e^-\to\Xi_{cc}+\bar{\Xi}_{cc}$ at the B factory in Table \ref{total cross section}. Four doubly charmed baryon masses $M_{\Xi_{cc}}=\{3.5,\;3.6,\;3.7,\;3.8\}$ GeV are adopted. We can see from the table that the cross section decreases as the mass increases. 

\begin{table}[ht]
    \caption{Cross sections (in units: ${\rm fb}$) for $e^++e^-\to\Xi_{cc}+\bar{\Xi}_{cc}$ at the B factories. Four doubly charmed baryon masses $M_{\Xi_{cc}}=\{3.5,\;3.6,\;3.7,\;3.8\}$ GeV are adopted}
    \begin{center}
    \renewcommand{\arraystretch}{1.5}
       \begin{tabular}{p{3cm}<{\centering} p{2.5cm}<{\centering} p{2.5cm} <{\centering} p{2.5cm}<{\centering} p{2.5cm} <{\centering}}
        \toprule
        \botrule
            $M_{\Xi_{cc}}$   & 3.5 GeV &  3.6 GeV & 3.7 GeV & 3.8 GeV  \\
        \hline
        $\sigma $ (fb) &
        $3.787 $ &
        $3.485$ &
        $3.193$ &
        $2.911$\\ 
        \botrule
      \end{tabular}
    \end{center}
    \label{total cross section}
\end{table}

Considering that the integrated luminosity at Belle II experiment is about ${\cal L}_{\mathrm{int}}=50 \; {\rm ab^{-1}}$ \cite{Belle-II:2018jsg}, and the BaBar experiment is designed to have an integrated luminosity of about ${\cal L}_{\mathrm{int}}=424.18 \; {\rm fb^{-1}}$ \cite{Polat:2024nhp,BaBar:2013agn} in their full running period, we present the total production event yields of doubly charmed baryons for Belle II and BaBar in Table \ref{event yields}. As can be seen from the table, the total yields of exclusive doubly charmed baryon production at Belle II exceed those at BaBar by more than two orders of magnitude, owing to the larger integrated luminosity. However, the detection prospects for the relevant processes at both Belle II and BaBar are unfavorable once detection efficiencies and decay branching fractions are taken into account.

\begin{table}[ht]
    \caption{Total yields for exclusive production for $e^++e^-\to\Xi_{cc}+\bar{\Xi}_{cc}$ at Belle II and BaBar. Four doubly charmed baryon masses $M_{\Xi_{cc}}=\{3.5,\;3.6,\;3.7,\;3.8\}$ GeV are adopted.}
    \begin{center}
    \renewcommand{\arraystretch}{1.3}
       \begin{tabular}{p{3cm}<{\centering} p{2.5cm}<{\centering} p{2.5cm} <{\centering} p{2.5cm}<{\centering} p{2.5cm} <{\centering}}
        \toprule
        \botrule
            $M_{\Xi_{cc}}$  & 3.5 GeV &  3.6 GeV & 3.7 GeV & 3.8 GeV  \\
        \hline
        $\text{Belle II} $ &
        $1.894 \times 10^5$ &
        $1.742 \times 10^5$ &
        $1.596 \times 10^5$ &
        $1.455 \times 10^5$\\ 
        \hline
        $\text{BaBar} $ &
        $1.607 \times 10^3$ &
        $1.478 \times 10^3$ &
        $1.354 \times 10^3$ &
        $1.235 \times 10^3$\\ 
        \botrule
      \end{tabular}
    \end{center}
    \label{event yields}
\end{table}

Finally, we consider the uncertainty from the non-perturbative long-distance matrix elements. As indicated by Eq. (\ref{sigma}), the cross section is proportional to $|\Psi_{cc[1S]}(0)|^4$. The wave functions at the origin $|\Psi_{cc[1S]}(0)|$ can be calculated under many kinds of potential models, for example K$^2$O potential \cite{Kiselev:2002iy} and Buchm\"uller-Tye (B.T.) potential \cite{Kiselev:2001fw}, which give the wave function values $|\Psi_{cc[1S]}(0)|^2_{K^2O}=0.0218 \ {\rm GeV^3}$ and $|\Psi_{cc[1S]}(0)|^2_{B.T.}=0.0224 \ {\rm GeV^3}$. Within the calculation framework adopting the K$^2$O potential, the total cross sections are $\{1.183,\;1.089,\;0.998,\;0.910\}$ fb for $M_{\Xi_{cc}}=\{3.5,\;3.6,\;3.7,\;3.8\}$ GeV, respectively.
All of these quantities decrease by approximately $68.75 \%$ compared to calculations using the Power-low potential.
For the B.T. potential model, the corresponding values are $\{1.249,\;1.150,\;1.054,\;0.961\}$ fb for $M_{\Xi_{cc}}=\{3.5,\;3.6,\;3.7,\;3.8\}$ GeV, respectively, and all cross sections decrease by approximately $67.01 \%$ compared to Power-low potential calculations. We may call attention to the fact that the long distance matrix elements in our calculation play an overall role, so if there are more accurate values of wave functions at the origin of doubly charmed baryons in the future, we can update our numerical results easily.

\subsection{Differential distributions}

\begin{figure}[htb]
\centering
\begin{minipage}{0.4\textwidth}
\includegraphics[width=\textwidth]{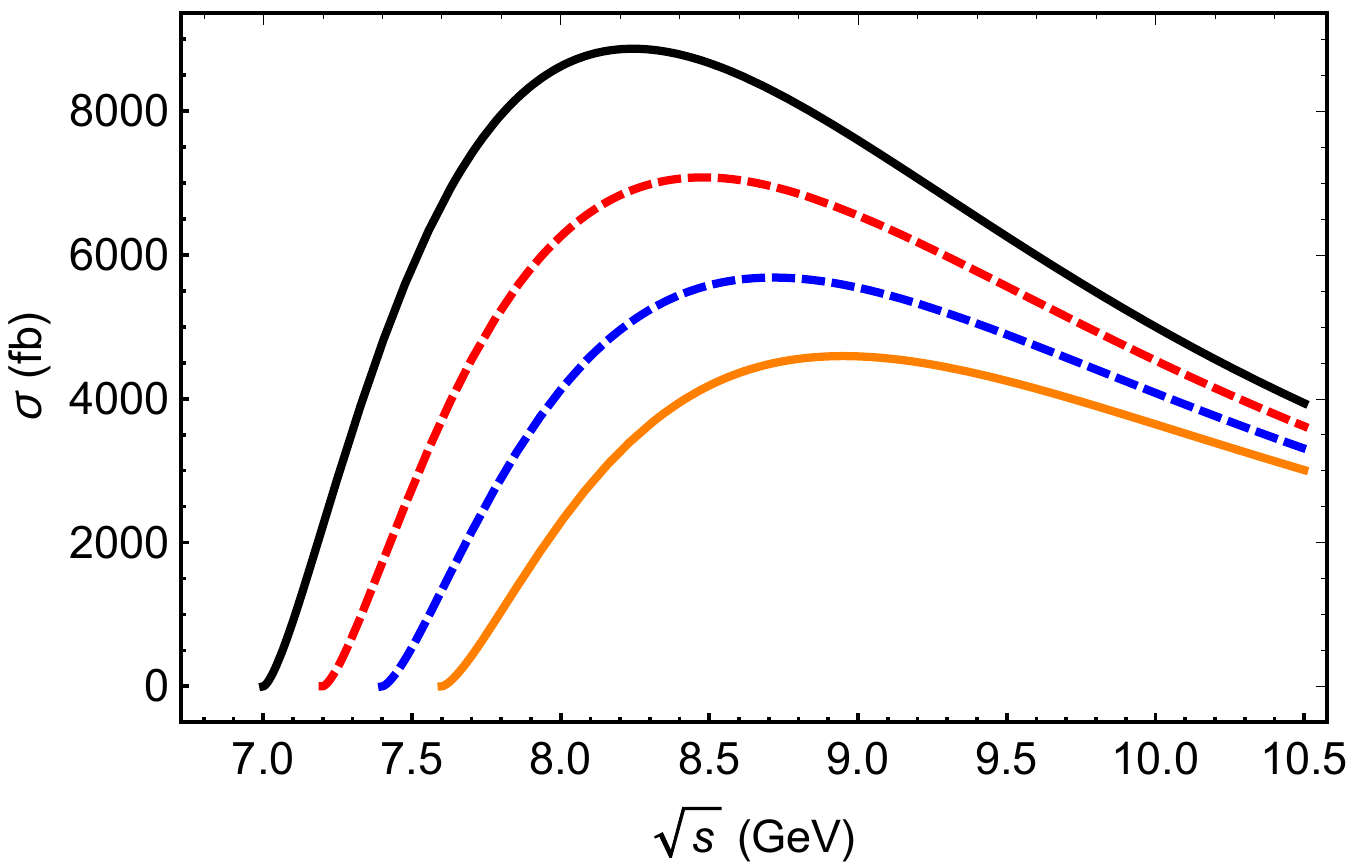}
\end{minipage}
\begin{minipage}{0.52\textwidth}
\includegraphics[width=\textwidth]{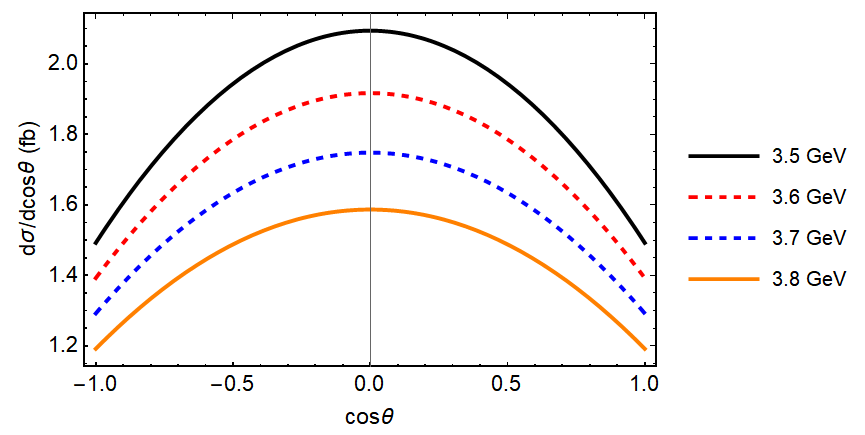}
\end{minipage}

\vspace{0.5em} 
\hspace{-0.1\textwidth}\begin{minipage}{0.4\textwidth}
\includegraphics[width=\textwidth]{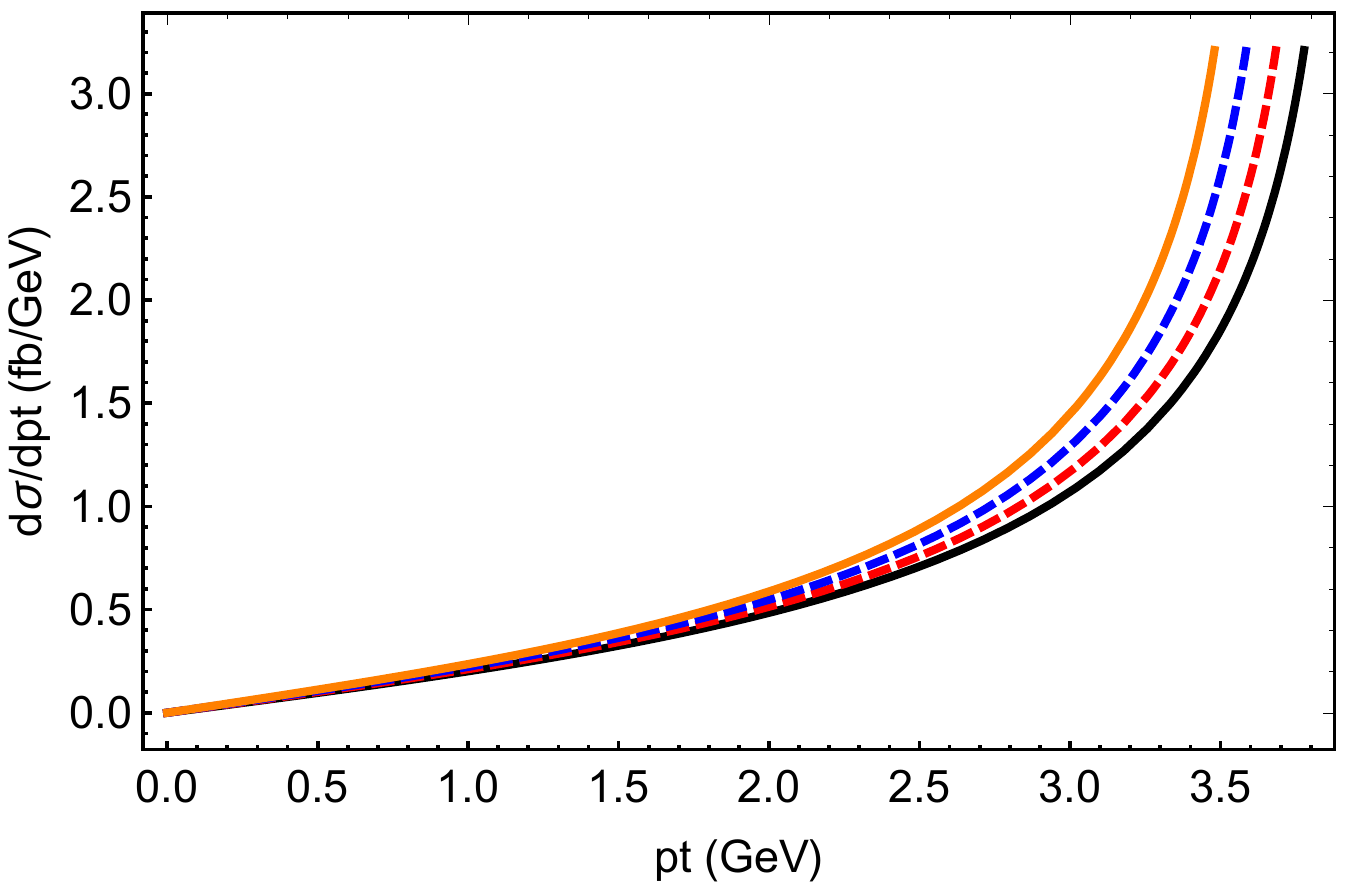}
\end{minipage}
\begin{minipage}{0.4\textwidth}
\includegraphics[width=\textwidth]{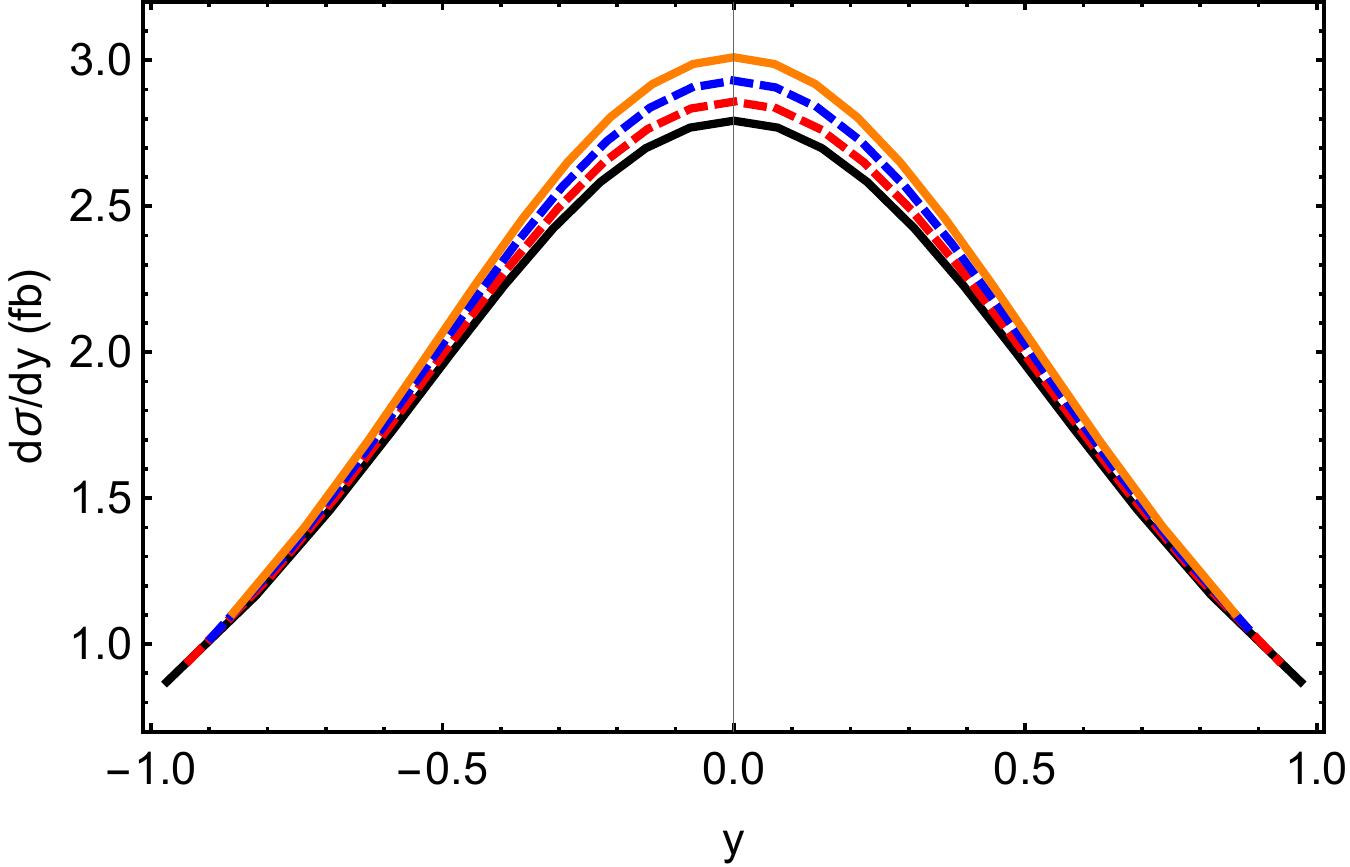}
\end{minipage}

\caption{Cross sections of exclusive production of doubly charmed baryons versus the CM energy $\sqrt{s}$, the differential angle distributions $d\sigma/dcos\theta$, the differential $p_t$ distributions $d\sigma/dp_t$ and the differential rapidity distributions $d\sigma/dy$ for $e^++e^-\to\Xi_{cc}+\bar{\Xi}_{cc}$. Four doubly charmed baryon masses $M_{\Xi_{cc}}=3.5$ GeV (black line), $3.6$ GeV (red dotted line), $3.7$ GeV (blue dotted line) and $3.8$ GeV (orange line) are adopted.} 
    \label{uncertainties}
\end{figure}

We next investigate the differential distributions for $e^++e^-\to\Xi_{cc}+\bar{\Xi}_{cc}$. In Fig. \ref{uncertainties} we display the cross sections versus the CM energy $\sqrt{s}$, the differential angle distributions $d\sigma/dcos\theta$, the differential transverse momentum distributions $d\sigma/dp_t$ and the differential rapidity distributions $d\sigma/dy$. 

As can be seen from the figure in the upper left corner, the production cross sections rise first and then decrease slowly with the increase of $\sqrt{s}$, reaching their peaks at around $\sqrt{s}=8.5 \; {\rm GeV}$. The differential angular distributions $d\sigma/dcos\theta$ (upper right) exhibit pronounced symmetric arch shapes. Here $\theta$ denotes the angle between the collision axis and the momentum of the final-state doubly charmed baryon or antibaryon.
For the transverse momentum distributions (bottom left), we can make use of angle distributions as follows,
\begin{eqnarray}
\frac{d\sigma}{dp_t}=\left|\frac{dcos\theta}{dp_t}\right| \left(\frac{d\sigma}{dcos\theta}\right)  =\frac{p_t}{|\vec{p}_1| \sqrt{|\vec{p}_1|^2-p^2_t}} \left(\frac{d\sigma}{dcos\theta}\right).
\label{eq:dpt}
\end{eqnarray}
As the figure shows, the differential cross sections versus $p_t$ are increasing monotonically.
By observing the figure of rapidity distribution (bottom right), We can observe a prominent symmetric bump in the central rapidity region.

\section{SUMMARY}
\label{sec:summary}

Within the NRQCD framework we study the exclusive processes of $\Xi_{cc} +\bar{\Xi}_{cc}$ through electron-positron fusion at B factory. The total cross sections are summarized in Table \ref{total cross section} for $M_{\Xi_{cc}}=\{3.5,\;3.6,\;3.7,\;3.8\}$ GeV. The total event yields are given in Table \ref{event yields}, and the uncertainties originating from the non‑perturbative long‑distance matrix elements are discussed. The various differential distributions are shown in Fig.~\ref{uncertainties}. We find that the detection prospects become unfavorable once the decay branching fractions and reconstruction efficiencies of final‑state particles are taken into account for both Belle II and BaBar.


\vspace{0.5cm} {\bf Acknowledgments}
This work is supported in part by the National Natural Science Foundation of China (12105162, 12305106, 12235008, 12321005, 12475083), the Natural Science Foundation of Shandong Province (ZR2021QA058, ZR2021QA040) and the Youth Innovation Technology Project of Higher School in Shandong Province (2023KJ146).



\end{document}